\documentclass[nohyper,12pt,letterpaper]{JHEP3}
\usepackage{epsfig}
\usepackage[latin1]{inputenc}
\usepackage{bbm,amsfonts}
\usepackage{graphicx}
\usepackage{amssymb,amsmath}

\author{Marco S. Bianchi$^\ast$,
  Marta Leoni$^{\dag, \hash}$,
  Matias Leoni$\ddag$,  
  Andrea Mauri$^{\dag, \hash}$,
  Silvia Penati$^{\flat}$,
  and Alberto Santambrogio$^{\hash}$\\\\
  $^\ast$Institut f\"ur Physik,
Humboldt-Universit\"at zu Berlin,
Newtonstra{\ss}e 15, 12489 Berlin, Germany \\
  $^\dag$Dipartimento di Fisica dell'Universit\`a degli studi di Milano, via Celoria 16, I-20133 Milano, Italy\\
  $^\hash$ INFN, Sezione di Milano, via Celoria 16, I-20133 Milano, Italy\\
  $^\ddag$Physics Department, FCEyN-UBA \& IFIBA-CONICET\
Ciudad Universitaria, Pabell\'on I, 1428, Buenos Aires, Argentina \\
  $^\flat$Dipartimento di Fisica, Universit\`a degli studi di Milano--Bicocca and
  INFN, Sezione di Milano--Bicocca, Piazza della Scienza 3, I-20126 Milano, Italy \\
  \qquad\\
  E-mail: \email{marco.bianchi@physik.hu-berlin.de, marta.leoni@mi.infn.it, 
  leoni@df.uba.ar, andrea.mauri@mi.infn.it,
    silvia.penati@mib.infn.it, alberto.santambrogio@mi.infn.it} }

\abstract{We evaluate ABJM observables at two loops, for any value of the rank $N$ of the gauge group. We compute the color subleading contributions to the four--point scattering amplitude in ABJM at two loops. Contrary to the four dimensional case,  IR divergent $N$--subleading contributions are proportional to leading poles in the regularization parameter. We then exploit the non--planar calculation for the amplitude to derive an expression for the two--loop Sudakov form factor at any $N$. In the planar limit the result coincides with the one recently obtained in literature by using Feynman diagrams and unitarity. Finally, we analyze the subleading contributions to the light--like four--cusps Wilson loop and interpret the result in terms of the non--abelian exponentiation theorem. All these perturbative results satisfy the uniform transcendentality principle, hinting at its validity in ABJM beyond the planar limit.}

\preprint{June 2013\\ HU-EP-13/29\\ IFUM-1012-FT}

\title{ABJM amplitudes and WL at finite $N$}

\keywords{Chern--Simons matter theories, scattering amplitudes, Wilson loops, Form Factors}

\csname @addtoreset\endcsname{equation}{section}

\def\bseq{\begin{subequation}}  
\def\eseq{\end{subequation}}
\def\bsea{\begin{subeqnarray}}  
\def\esea{\end{subeqnarray}}

\newcommand{\beq}{\begin{equation}}
\newcommand{\bea}{\begin{eqnarray}}
\newcommand{\eea}{\end{eqnarray}}
\newcommand{\eeq}{\end{equation}}

\newcommand {\non}{\nonumber}

\renewcommand{\a}{\alpha}

\renewcommand{\d}{\delta}

\newcommand{\g}{\gamma}
\newcommand{\G}{\Gamma}

\newcommand{\e}{\epsilon}

\newcommand{\m}{\mu}

\newcommand{\F}{\Phi}

\newcommand{\Db}{\overline{D}}

\def\Mb{\kern 2pt\mathchoice
        {
         \vbox{\hrule width10pt height 0.4pt depth 0pt
         \kern 1.2pt\hbox{\kern -2pt$\displaystyle M$}}}
        {
         \vbox{\hrule width10pt height 0.4pt depth 0pt
         \kern 1.2pt\hbox{\kern -2pt$\textstyle M$}}}
        {
\vbox{\hrule width6pt height 0.4pt depth 0pt
         \kern 1.0pt\hbox{\kern -2pt$\scriptstyle M$}}}
        {
         \vbox{\hrule width5pt height 0.4pt depth 0pt
         \kern 0.8pt\hbox{\kern -2pt$\scriptscriptstyle M$}}}}

\def\Sb{\kern 2pt\mathchoice
        {
         \vbox{\hrule width6pt height 0.4pt depth 0pt
         \kern 1.2pt\hbox{\kern -2pt$\displaystyle S$}}}
        {
         \vbox{\hrule width6pt height 0.4pt depth 0pt
         \kern 1.2pt\hbox{\kern -2pt$\textstyle S$}}}
        {
         \vbox{\hrule width3.5pt height 0.4pt depth 0pt
         \kern 1.0pt\hbox{\kern -2pt$\scriptstyle S$}}}
        {
         \vbox{\hrule width3pt height 0.4pt depth 0pt
         \kern 0.8pt\hbox{\kern -2pt$\scriptscriptstyle S$}}}}

\def\Rb{\kern 2pt\mathchoice
        {
         \vbox{\hrule width5.5pt height 0.4pt depth 0pt
         \kern 1.2pt\hbox{\kern -2.5pt$\displaystyle R$}}}
        {
         \vbox{\hrule width5.5pt height 0.4pt depth 0pt
         \kern 1.2pt\hbox{\kern -2.5pt$\textstyle R$}}}
        {
         \vbox{\hrule width3.5pt height 0.4pt depth 0pt
         \kern 1.0pt\hbox{\kern -2.2pt$\scriptstyle R$}}}
        {
         \vbox{\hrule width3pt height 0.4pt depth 0pt
         \kern 0.8pt\hbox{\kern -2.2pt$\scriptscriptstyle R$}}}}

  \def\pp{{\mathchoice
      {
          \kern 1pt%
          \raise 1pt
          \vbox{\hrule width5pt height0.4pt depth0pt
            \kern -2pt
            \hbox{\kern 2.3pt
              \vrule width0.4pt height6pt depth0pt
              }
            \kern -2pt
            \hrule width5pt height0.4pt depth0pt}%
            \kern 1pt
       }
        {
          \kern 1pt%
          \raise 1pt
          \vbox{\hrule width4.3pt height0.4pt depth0pt
            \kern -1.8pt
            \hbox{\kern 1.95pt
              \vrule width0.4pt height5.4pt depth0pt
              }
            \kern -1.8pt
            \hrule width4.3pt height0.4pt depth0pt}%
            \kern 1pt
        }
        {
          \kern 0.5pt%
          \raise 1pt
          \vbox{\hrule width4.0pt height0.3pt depth0pt
            \kern -1.9pt  
            \hbox{\kern 1.85pt
              \vrule width0.3pt height5.7pt depth0pt
              }
            \kern -1.9pt
            \hrule width4.0pt height0.3pt depth0pt}%
            \kern 0.5pt
        }
        {
          \kern 0.5pt%
          \raise 1pt
          \vbox{\hrule width3.6pt height0.3pt depth0pt
            \kern -1.5pt
            \hbox{\kern 1.65pt
              \vrule width0.3pt height4.5pt depth0pt
              }
            \kern -1.5pt
            \hrule width3.6pt height0.3pt depth0pt}%
            \kern 0.5pt
        }
    }}

  \def\mm{{\mathchoice
               {
                 \kern 1pt
           \raise 1pt    \vbox{\hrule width5pt height0.4pt depth0pt
                  \kern 2pt
                  \hrule width5pt height0.4pt depth0pt}
                 \kern 1pt}
               {
                \kern 1pt
           \raise 1pt \vbox{\hrule width4.3pt height0.4pt depth0pt
                  \kern 1.8pt
                  \hrule width4.3pt height0.4pt depth0pt}
                 \kern 1pt}
               {
                \kern 0.5pt
           \raise 1pt
                \vbox{\hrule width4.0pt height0.3pt depth0pt
                  \kern 1.9pt
                  \hrule width4.0pt height0.3pt depth0pt}
                \kern 1pt}
               {
               \kern 0.5pt
         \raise 1pt  \vbox{\hrule width3.6pt height0.3pt depth0pt
                  \kern 1.5pt
                  \hrule width3.6pt height0.3pt depth0pt}
               \kern 0.5pt}
               }}

\def\pd{{\kern0.5pt
           + \kern-5.05pt \raise5.8pt\hbox{$\textstyle.$}\kern
0.5pt}}

\def\pmd{{\kern0.5pt
          \pm \kern-5.05pt
\raise6.3pt\hbox{$\textstyle.$}\kern1.5pt}}

\def\md{{\mathchoice
   {
      {{\kern 1pt - \kern-6.2pt \raise5pt\hbox{$\textstyle.$}\kern
1pt}}}
    {
      {{\kern 1pt - \kern-6.2pt \raise5pt\hbox{$\textstyle.$}\kern
1pt}}}
    {
      {\kern0.5pt - \kern-5.05pt
\raise3.4pt\hbox{$\textstyle.$}\kern0.5pt}}
    {
      {\kern0.5pt - \kern-5.05pt
\raise3.4pt\hbox{$\textstyle.$}\kern0.5pt}}}}

\def\beq{\begin{equation}}
\def\eeq{\end{equation}}
\def\bea{\begin{eqnarray}}
\def\eea{\end{eqnarray}}
\def\Tr{\mathrm{Tr}}

\def\a{\alpha}

\def\g{\gamma}

\def\d{\delta}
\def\e{\epsilon}

\def\th{\theta}

\def\G{\Gamma}

\def\F{\mathcal{F}}

\begin{document}

\section{Introduction}

In the past few years much progress has been achieved in the perturbative analysis of three dimensional Chern--Simons matter theories and especially of the ABJ(M) models \cite{ABJM, ABJ}.  Beside its independent relevance in the context of AdS/CFT correspondence, the three dimensional setup proves to be a good playground to check whether the mathematical  structures exhibited by $\mathcal{N}=4$ SYM have a counterpart in models which are {\it a priori} different in nature. 

A striking example is provided by the emergence of integrable structures in the spectral problem of the ABJM theory, which has been formulated along the lines of the $\mathcal{N}=4$ SYM case and then extensively checked (see \cite{Beisert:2010jr} for a review). With respect to integrability, the ABJM model looks surprisingly similar to $\mathcal{N}=4$ SYM  and independent features only become relevant at high perturbative orders \cite{Gromov:2008qe, Mauri:2013vd}.  

An indirect way to test the appearance of integrable structures is to study the on--shell sector of the theory. In the four--dimensional case a great effort has been devoted to the evaluation of scattering amplitudes, Wilson loops and form factors. These quantities have become important also in the context of AdS/CFT correspondence due to a number of remarkable stringy inspired properties they have been shown to possess.  On--shell scattering amplitudes exhibit a duality with light--like Wilson loops \cite{Drummond:2007aua}--\cite{Anastasiou:2009kna} and correlators of BPS operators \cite{Eden:2010zz},  exponentiation \cite{BDS}, enhanced dynamical symmetries like dual conformal \cite{Drummond:2006rz,Drummond:2008vq,Brandhuber:2008pf} and Yangian invariance \cite{DHP} and color/kinematics duality \cite{BCJ}.  

Subsequently, following the seminal work of Van Neerven \cite{vanNeerven:1985ja},  the perturbative computation of form factors and their supersymmetric extensions have been performed in $\mathcal{N}=4$ SYM up to three--loop order \cite{Bork:2010wf}--\cite{Bork:2012tt}.  Form factors have also been studied at strong coupling \cite{Alday:2007he, Maldacena:2010kp, Gao:2013dza} and conjectured to be dual to light--like periodic Wilson loops \cite{Alday:2007he, Maldacena:2010kp, Brandhuber:2010ad}. The existence of color/kinematics duality for form factors has also been proposed and verified in two and three--loop examples \cite{Boels:2012ew}.

In this respect the three--dimensional picture seems to slightly depart from the four dimensional case and, while a partial parallelism can still be traced, a precise definition of the above dualities requires more care.  
First, the tree--level  four and six--point amplitudes have been found explicitly and Yangian invariance has been established \cite{BLM, Huang:2010rn, HL} for all point amplitudes with the help of a three--dimensional form of BCFW recursion relations \cite{Gang:2010gy}. 

At loop level explicit computations are available for the four and six--point case up to two loops. 
The four--point one--loop complete superamplitude is of $\mathcal{O}(\epsilon)$ in dimensional regularization \cite{ABM,CH,BLMPS1}. At two loops  the planar amplitude can be written as a sum of dual conformal invariant integrals and has been found to coincide with the second order expansion of a light--like four--polygon Wilson loop \cite{CH,BLMPS1}. This points to the fact that a  Wilson loop/scattering amplitude duality might exist even if a strong coupling interpretation of the duality is less straightforward \cite{ADO}--\cite{Colgain:2012ca} with respect to the four dimensional case \cite{BM,Beisert:2008iq}. Moreover,  the two--loop four--point ABJM amplitude  surprisingly  matches the amplitude of $\mathcal{N} = 4$ SYM theory at one loop \cite{CH,BLMPS1}. 

Beyond four points the connection with the four--dimensional case gets looser. In the ABJM model all the odd legs amplitudes are forced to be vanishing by gauge invariance. The six--point one--loop amplitude has been shown not to vanish \cite{Bargheer:2012cp,Bianchi:2012cq} contrary to the hexagon light--like Wilson loop \cite{HPW, BLMPRS}. This suggests that if a WL/amplitude duality exists it must  be implemented with a proper definition of a (super)Wilson loop. At two loops the six--point amplitude has been computed analytically in \cite{CaronHuot:2012hr} and shown to exhibit some similarity with the one--loop MHV amplitude in $\mathcal{N}=4$ SYM, even if the identification experienced for the four--point amplitude gets spoiled. 

Very recently, an analysis of the form factors has been initiated also for the ABJM model, where  computations for BPS operators have been performed through unitarity cuts \cite{Brandhuber:2013gda} and component Feynman diagrams formalism \cite{Young:2013hda}.

All the remarkable properties detailed above have been found to hold in the large $N$ limit of $\mathcal{N}=4$ SYM and ABJM, which seems to be the regime where interaction simplifies in such a way  that dualities and integrability can occur.  Nevertheless, it is interesting to look at what happens to the subleading corrections of Wilson loops, amplitudes and form factors. 

For scattering amplitudes in four dimensions their complete evaluation including subleading partial amplitudes is constrained by underlying BCJ relations \cite{BCJ}. These in turn are useful for determining gravity amplitudes as a double copy \cite{Bern:2007hh,Bern:2010ue,Bern:2010yg}. Moreover, interesting relations between the IR divergences of subleading ${\cal N}=4$ SYM amplitudes and ${\cal N}=8$ supergravity ones have been pointed out \cite{Naculich:2008ys,Naculich:2011my,Naculich:2013xa}. In three dimensions a proposal for BCJ like relations governed by a three algebra structure has been suggested \cite{Bargheer:2012gv}. Despite checks at tree level \cite{Huang:2012wr}, it would be interesting to understand how it applies to loop amplitudes. This, in fact, requires their knowledge at finite $N$.

Furthermore, the three dimensional BLG theory \cite{Bagger:2006sk}--\cite{Gustavsson:2007vu}, possessing $OSp(4|8)$ superconformal invariance is realized as a $SU(2)\times SU(2)$ theory \cite{VanRaamsdonk:2008ft}, where the planar limit cannot be taken. Therefore inspection of BLG amplitudes at loop level inevitably requires working at finite $N$.

It is the purpose of this paper to tackle the computation of the four--point scattering amplitude, the four--cusped light--like Wilson loop and the form factor up to two loops at finite $N$.

For the ABJM scattering amplitude in Section \ref{sec:amplitude}, we use the supergraph approach of \cite{BLMPS1,BLMPS2}, which makes possible to complete the computation in terms of a limited number of Feynman diagrams. To get the subleading contributions it is necessary to add a new non--planar diagram, leading to a non--trivial non--planar integral which we solve explicitly in Appendix B. The result for the complete four--point amplitude is given in eq. (\ref{result}). As a by--product, we also provide the full four--point scattering amplitude ratio of BLG theory, see eq. (\ref{resultBLG}). 

Infrared divergences also appear in the subleading--in--$N$ contributions as poles in the dimensional regularization parameter. However, in contrast with the ${\cal N}=4$ SYM amplitude where subleading terms have milder divergences, the three dimensional amplitude exhibits a uniform leading $\e^{-2}$ pole, both in the leading and subleading parts. As we argue in the main text, this can be understood as a consequence of the different color structures underlying amplitudes in the two cases.

In Section \ref{sec:formfactor} we use the information collected in the computation of the amplitude to evaluate the bilinear Sudakov form factor at any value of $N$. Indeed, it turns out that the superspace computation of this object can be reduced to a sum of $s$--channel contributions given by a subset of diagrams involved in the amplitude, albeit with different color factors. Taking the planar limit our result matches the form factor given in \cite{Brandhuber:2013gda, Young:2013hda} and extend it to finite $N$ for both ABJM and ABJ theory. 

Finally, in Section \ref{sec:WL} we give an expression for the two--loop subleading contributions to the light--like Wilson loop with four cusps. Subleading corrections emerge only in the pure Chern--Simons sector and give rise to a simple structure, which can be understood as a result of the non--abelian exponentiation property of Wilson loops \cite{Gatheral:1983cz,Frenkel:1984pz} and the vanishing of the one--loop contribution \cite{HPW,BLMPRS}.

All the results we obtain exhibit uniform transcendentality two. This suggests that the {\em maximal transcendentality principle} \cite{Kotikov:2001sc,Kotikov:2004er} likely applies to ABJM at finite $N$.

\section{Subleading contributions to the four--point amplitude}\label{sec:amplitude}

As described in details in \cite{BLMPS1,BLMPS2}, the evaluation of the four--point amplitude in $U(N) \times U(N)$ ABJM theory at two loops is doable by a direct supergraph approach. 

The preliminary observation that allows to simplify the calculation is that all ABJM amplitudes with four external particles are related by simple supersymmetric Ward identities \cite{ABM}. As a consequence, the two--loop result for a particular component, divided by its tree level counterpart, is sufficient for reconstructing the whole superamplitude at that order. 

It is then convenient to focus on a particular configuration of external particles for which a limited number of diagrams enters the calculation. 
In ${\cal N}=2$ superspace formulation\footnote{See Appendix A for notations and conventions on the ABJM theory in ${\cal N}=2$ superspace.} this is the case for chiral amplitudes, which can be easily read from quantum corrections to the superpotential once the external fields are set on--shell.  
 
\FIGURE{
    \centering
    \includegraphics[width=0.7\textwidth]{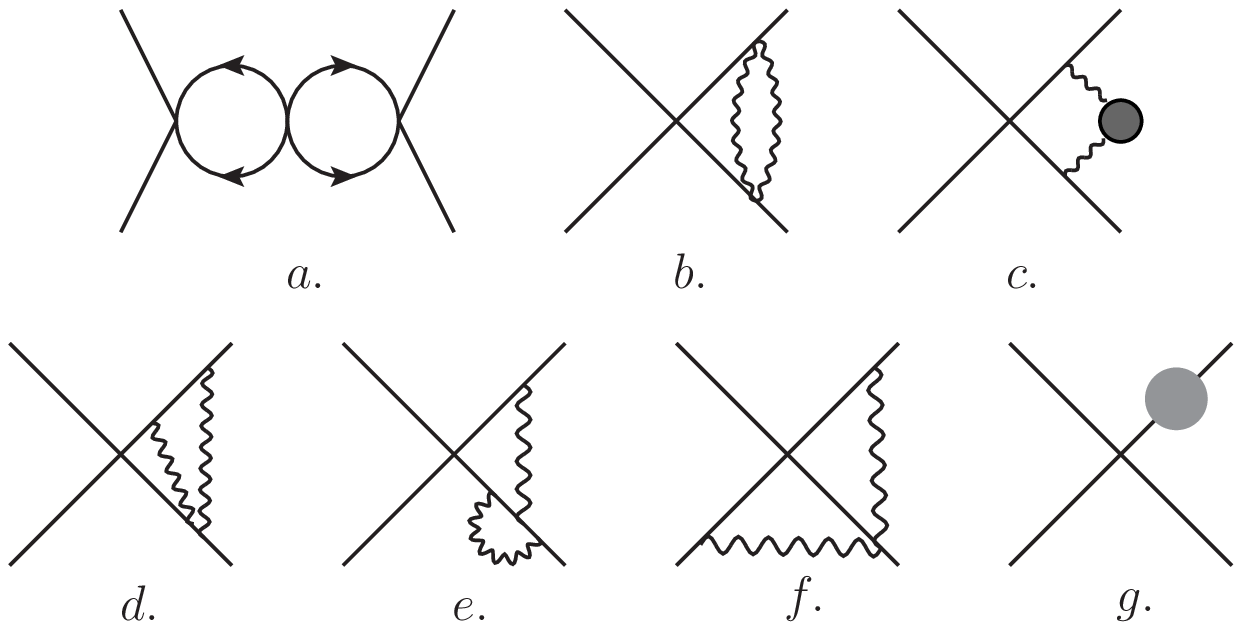}
    \caption{Planar diagrams contributing to the two--loop four--point scattering amplitude. The dark--gray blob represents one--loop corrections and the light--gray blob two--loop ones.}
    \label{fig:1set}
}

In \cite{BLMPS1,BLMPS2} this program has been carried out in the planar limit.  While the one--loop contribution vanishes, non--trivial two--loop corrections are given by supergraphs depicted in Fig. \ref{fig:1set}, for both terms appearing in the superpotential (\ref{action}). Wavy lines represent the gauge superfields of the two $U(N)$'s. The sum  over all possible configurations of $V$ and $\hat{V}$ has to be understood.    

In order to compute the complete amplitude we perform D--algebra on the graphs to reduce them to local expressions in superspace, proportional to ordinary loop integrals. These integrals are generally divergent and we deal with them by dimensional regularization, $d=3-2\epsilon$. Their explicit evaluation has been presented in \cite{BLMPS1,BLMPS2}. 

The result for the planar four--point amplitude divided by its tree level counterpart\footnote{Tree level amplitudes are simply given by the factor in front of the superpotential and correspond to single trace partial amplitudes.}, up to ${\cal O}(\epsilon)$ terms reads
\begin{equation}
\label{planar}
{\cal M}_4^{planar} \equiv \frac{{\cal A}_4^{(2)}|_{planar}}{{\cal A}_4^{(0)}} = \left(\frac{N}{K}\right)^2 \left(-\frac{(s/\m'^2)^{-2 \e}+(t/\m'^2)^{-2 \e}}{(2\e)^2} + \frac{1}{2} \log^2 \frac{s}{t} + 4\zeta_2
 + 3 \log^2 2\right)
\end{equation}
where $s,t$ are the Mandelstam variables and $\m'^2= 8 \pi e^{-\g_E}\m^2$ with $\m^2$ the IR scale of dimensional regularization.

This result exhibits very interesting properties. First of all, it can be obtained as a combination of dual conformally invariant integrals, as shown by a generalized unitarity computation \cite{CH}. It matches the form of the two--loop correction to the four--cusped light--like Wilson loop, hinting at a possible Wilson loop/amplitude duality in ABJM.
Finally the expression of the two--loop ABJM amplitude ratio is exactly the same as the one--loop one in ${\cal N}=4$ SYM upon rescaling the dimensional regularization parameter $\epsilon$ and to order ${\cal O}(\epsilon)$.
This relation has been sharpened in \cite{BLP}, where an all--order in $\epsilon$ identity has been derived between the two objects.
The similarity with the ${\cal N}=4$ SYM amplitude and the fact that they obey the same anomalous conformal Ward identities suggests that the ABJM four--point amplitude can also exponentiate \cite{CH,BLMPS1,BLP}.
It has to be stressed that the Wilson loop/amplitude duality and dual conformal invariance are not supported at strong coupling by AdS/CFT arguments \cite{AM,Beisert:2008iq}, as it is not clear whether fermionic T--duality could be a symmetry of the dual string sigma model \cite{ADO}--\cite{Colgain:2012ca}.
Moreover, the duality with the bosonic Wilson loop doesn't extend beyond four points since $n$--point amplitudes are no longer MHV for $n \geq 6$.\\

\FIGURE{
    \centering 
    \includegraphics[width=0.2\textwidth]{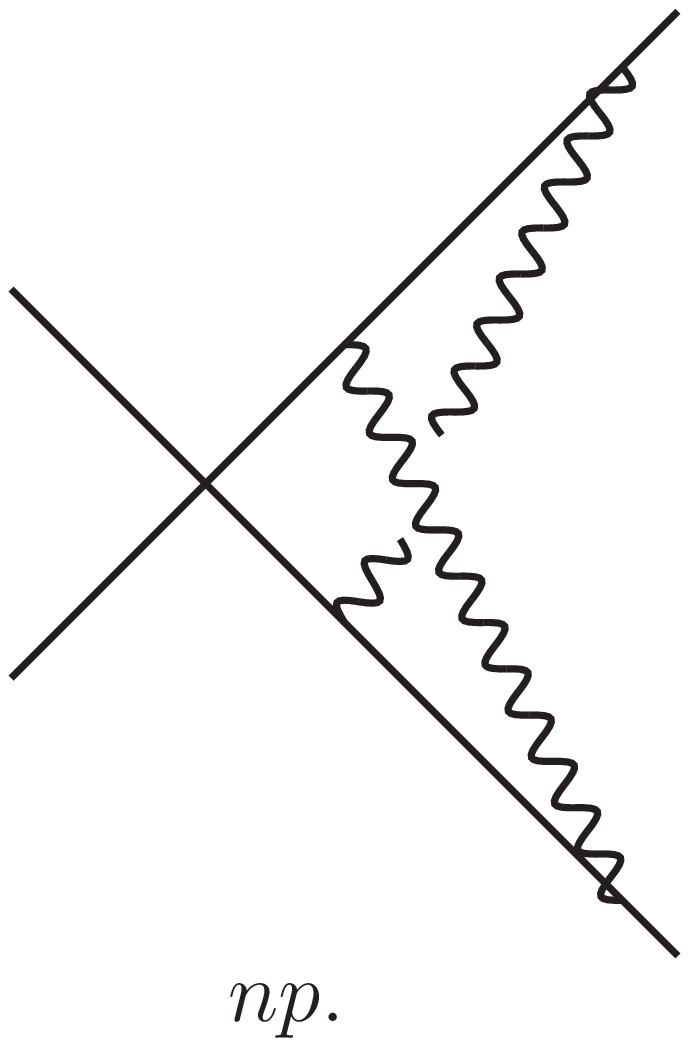}
    \caption{Nonplanar diagram contributing to the two--loop four--point scattering amplitude.}
    \label{fig:np} 
}

We now extend the previous result to the case of finite $N$. This requires taking into account contributions from non--planar diagrams plus $N$--subleading terms in the color factor associated to each diagram. Performing a preliminary color decomposition of the amplitude,  we expect single trace contributions appearing with a leading $N^2$ and a subleading $N^0$ behaviour, and double trace contributions with a subleading $N$ behaviour.

As in the planar case, at one loop it turns out that the momentum integrals, when evaluated in dimensional regularization are ${\cal O} (\e)$ and therefore negligible. 
 
At two loops,  we have to consider the non--planar version of diagrams in Fig. \ref{fig:1set} plus new genuinely non--planar graphs. In fact, it turns out that non--vanishing contributions only come from one new non--planar topology, depicted in Fig. \ref{fig:np}. 

Evaluating the complete color structure of each diagram in Fig. \ref{fig:1set} it turns out that double trace terms always cancel. 
Therefore, taking into account all channels and exploiting the results for the loop integrals in \cite{BLMPS1,BLMPS2} we can list the contribution from each diagram to the single trace partial amplitude, divided by its tree level counterpart
\begin{eqnarray} 
{\cal M}^{(a)} &=& \  \left(\frac{4\pi N}{K}\right)^2\ \left({\cal D}_a(s)+{\cal D}_a(t) \right)  + 4 \ \left(\frac{4\pi}{K}\right)^2\  {\cal D}_a(u) \nonumber \\
{\cal M}^{(b)} &=& \ \frac12\ \left(\frac{4\pi N}{K}\right)^2 \ \left({\cal D}_b(s)+{\cal D}_b(t)\right) +  \left(\frac{4\pi}{K}\right)^2 \left({\cal D}_b(s)+{\cal D}_b(t)+3\ {\cal D}_b(u)\right) \nonumber \\
{\cal M}^{(c)} &=& - \frac32 \left(\frac{4\pi N}{K}\right)^2 \left({\cal D}_b(s)+{\cal D}_b(t)-2 {\cal G}_d^p\right) + 
3 \left(\frac{4\pi}{K}\right)^2 \left({\cal D}_b(s)+{\cal D}_b(t)-{\cal D}_b(u)-{\cal G}_d^p \right) \nonumber \\
{\cal M}^{(d)} &=& \ 2\ \left(\frac{4\pi N}{K}\right)^2 \ \left({\cal D}_d(s)+{\cal D}_d(t)\right) + 4\left(\frac{4\pi}{K}\right)^2 \left({\cal D}_d(s)+{\cal D}_d(t)+3\ {\cal D}_d(u)\right) \nonumber \\
{\cal M}^{(e)} &=& \ 4\ \left(\frac{4\pi N}{K}\right)^2 \  \left({\cal D}_e(s)+{\cal D}_e(t)\right) - 8 \left(\frac{4\pi}{K}\right)^2 \left({\cal D}_e(s)+{\cal D}_e(t)-{\cal D}_e(u)\right) \nonumber \\
{\cal M}^{(f)} &=& \  \left(\frac NK\right)^2  \left(\frac{1}{2}\log^2 \frac st +3\ \zeta_2 \right) + \frac{1}{K^2} \left(-\log^2 \frac st+\log^2 \frac us+\log^2\frac tu+6\ \zeta_2 \right) + \mathcal{O}(\epsilon) \nonumber  \\
\label{planar2}
{\cal M}^{(g)} &=& \ -3 \left(\frac{4\pi N}{K}\right)^2 {\cal G}_{d}^p\  + 3 \left(\frac{4\pi}{K}\right)^2  {\cal G}_{d}^p 
\end{eqnarray} 
The loop integrals ${\cal D}$ have been defined in \cite{BLMPS1,BLMPS2}. Upon evaluation, they are given by 
\begin{align}
\label{integrals1}
 {\cal D}_a(s) &=  - G[1,1]^2  (s/\m^2)^{-2\e}  \nonumber \\
 {\cal D}_b(s) &= \frac{2 G[1,1]\Gamma(1+2\e)\Gamma^2(-2\e)}{(4\pi)^{d/2}\Gamma(1/2-3\e)} (s/\m^2)^{-2\e} \nonumber \\
 {\cal D}_d(s) &= -\frac{\Gamma^3 \left(\frac12-\e \right) \Gamma^2(-2 \e) \Gamma(2\e+1)}{(4 \pi )^{d} \Gamma \left(\frac12-3 \e \right) \Gamma^2 (1-2 \e)} (s/\m^2)^{-2\e}  \nonumber \\
 {\cal D}_e(s) &= {\cal D}_d(s)+ \frac12 {\cal D}_b(s) 
\end{align}
where $G[a,b]$ is defined in eq. (\ref{formula}). The  integral ${\cal G}_d^p$ is the linear combination 
\beq
{\cal G}_{d}^p = {\cal G}_d(p_1) + {\cal G}_d(p_2) + {\cal G}_d(p_3) + {\cal G}_d(p_4) 
\eeq
where ${\cal G}_d$ is on--shell vanishing, but otherwise IR divergent
\beq
\label{integral2}
{\cal G}_d(p_1) = \frac{G[1,1] G[1, 3/2 +\e]}{(p_1^2)^{2\e}} 
\eeq
It gets cancelled between diagrams $(c)$ and $(g)$.

For the non--planar contribution of Fig. \ref{fig:np} we still experience the cancellation of double trace contributions, once we take into account all possible configurations of gauge vector superfields. The contribution to the partial amplitude divided by the tree level counterpart reads
\begin{eqnarray}\label{eq:non--planar}
 {\cal M}^{(np)} &=& \ 4\, \left(\frac{4\pi}{K}\right)^2 \left( {\cal D}_{np}(s) + {\cal D}_{np}(t) + {\cal D}_{np}(u) \right)
\end{eqnarray}
where the non--planar Feynman integral
\begin{equation}
{\cal D}_{np}(s) = -(\mu^2)^{2\e} \int \frac{d^d k}{(2\pi)^d} \frac{d^d l}{(2\pi)^d} \frac{ \Tr ( (k+l)\ k\ l \ (k+l)\ p_4\ p_3 )} {k^2 (k+l-p_3)^2 (k+p_4)^2 (l-p_3)^2 \ (k+l+p_4)^2 \ l^2}
\end{equation}
is solved in Appendix \ref{app:integral}. Taking its leading contributions in the $\e$--expansion we obtain
\begin{equation}
\label{np}
{\cal D}_{np}(s) =  \frac{e^{-2\e \g_E} (16 \pi)^{2 \epsilon} (s/\mu^2)^{-2 \epsilon}}{64 \pi^2} \left(\frac{1}{(2\epsilon)^2}-\frac{\pi ^2}{24}-4 \log ^2 2\right)
\end{equation}
Now, summing these terms to the ones in eq. (\ref{planar2}) suitably expanded in powers of $\e$, it is easy to see that simple poles can be reabsorbed into a redefinition of  the regularization mass parameter $\mu'^{2}= 8 \pi e^{-\gamma_E}\mu^2$, which is the same for both planar and non--planar contributions.  The complete four--point amplitude at two loops then reads up to $\mathcal{O}(\epsilon)$
\begin{align}
\label{result}
{\cal M}_4 &= \left(\frac{N}{K}\right)^2 \left(-\frac{(s/\m'^2)^{-2 \epsilon }+(t/\m'^2)^{-2 \epsilon }}{(2 \epsilon)^2} + \frac{1}{2} \log ^2 \frac{s}{t} + \frac{2 \pi ^2}{3} + 3 \log^2 2\right) + \\ \nonumber
& ~ + \frac{1}{K^2} \left(\frac{2\, (s/\m'^2)^{-2 \epsilon }+2\, (t/\m'^2)^{-2 \epsilon }-2\, (u/\m'^2)^{-2 \epsilon }}{(2\epsilon) ^2} + 2\, \log \frac{s}{u}\, \log \frac{t}{u} + \frac{\pi ^2}{3} - 3 \log ^2 2\right)
\end{align}
We note that the result exhibits uniform transcendentality. This is a first indication that the maximal transcendentality principle \cite{Kotikov:2001sc,Kotikov:2004er} could apply to ABJM amplitudes beyond the planar limit, as is the case in ${\cal N}=4$ SYM in four dimensions \cite{Naculich:2008ys}.

We emphasize once again that along the calculation all double trace contributions have cancelled separately for each single topology, leading to a final result for the amplitude which contains only single trace terms. The technical reason for such a pattern can be traced back to the fact that, at least at this order, double trace contributions from diagrams with one gauge vector get compensated by analogous contributions where the other gauge vector runs inside the loops. 

Even if the disappearance of double trace structures from the final result was not {\em a priori} expected, it has a simple interpretation in terms of unitarity, as we are now going to discuss in detail. 

When constructing the whole two--loop four point amplitude from unitarity cuts, we have to take into account all two--particle cuts separating it into a one--loop and a tree level four--point amplitudes, as well as three--particle cuts dividing it into two five--point tree level amplitudes. Since the latter vanish, color structures do not emerge from three--particle cuts and we can focus on two--particle ones.
We can concentrate for instance on the two--particle cut in the $s$--channel separating the amplitude into a one loop ${\cal A}^{(1)}_4(1,2,A,B)$ and a tree level ${\cal A}^{(0)}_4(B,A,3,4)$ four--points.

In the color space and at any order in loops four--point amplitudes of the form $( (A^i)^{i_1}_{\ \bar{i}_1} (B_j)^{\bar{i}_2}_{\ i_2}  (A^k)^{i_3}_{\ \bar{i}_3} (B_l)^{\bar{i}_4}_{\ i_4})$ can be expanded on a basis of four independent structures, two single and two double traces (we remind that matter fields are in the bifundamental representation of the gauge group)
\bea
\label{traces}
&& [1,2,3,4] = \d^{\bar{i}_2}_{\bar{i}_1} \, \d^{i_3}_{i_2}  \, \d^{\bar{i}_4}_{\bar{i}_3} \, \d^{i_1}_{i_4} \quad  , \quad 
[1,4,3,2] =  \d^{\bar{i}_4}_{\bar{i}_1} \, \d^{i_3}_{i_4}  \, \d^{\bar{i}_2}_{\bar{i}_3} \, \d^{i_1}_{i_2} \non \\
&& [1,2][3,4] =  \d^{i_1}_{i_2} \, \d^{\bar{i}_2}_{\bar{i}_1}  \, \d^{i_3}_{i_4} \, \d^{\bar{i}_4}_{\bar{i}_3} \quad , \quad 
[1,4][3,2] =   \d^{i_1}_{i_4} \, \d^{\bar{i}_4}_{\bar{i}_1}  \, \d^{i_3}_{i_2} \, \d^{\bar{i}_2}_{\bar{i}_3} 
\eea
Using the results of \cite{Brandhuber:2013gda} we see that at one loop ${\cal A}^{(1)}_4(1,2,A,B)$ contains three possible structures 
\begin{equation}
[1,2,A,B] + [1,B,A,2] \quad,\quad [1,2][A,B] \quad {\rm and} \quad [1,B][A,2]
\end{equation}
whereas the tree level amplitude ${\cal A}^{(0)}_4(B,A,3,4)$ enters with
\begin{equation}
[3,4,B,A] - [3,A,B,4]
\end{equation}
Combining the traces from the two lower order amplitudes to obtain the two--loop structure, for the selected channel we find  
\begin{align}
N\, \left( [1,2,A,B] + [1,B,A,2] \right) \times \left( [3,4,B,A] - [3,A,B,4] \right) = \nonumber\\= N^2 [1,2,3,4] - N [1,2][3,4] + N[1,2][3,4] - N^2 [1,4,3,2] \, ,
\end{align}
\begin{equation}
[1,2][A,B] \times \left( [3,4,B,A] - [3,A,B,4] \right) = N\,[1,2][3,4] - N\,[1,2][3,4]
\end{equation}
and
\begin{equation}
[1,B][A,2] \times \left( [3,4,B,A] - [3,A,B,4] \right) = [1,2,3,4] - [1,4,3,2]
\end{equation}
From these relations it is easy to see that all double traces cancel. Repeating the same analysis for all channels one can prove the absence of double traces in the two--loop amplitude.

\subsection{IR divergences}

The evaluation of the four--point amplitude at finite $N$ reveals that IR divergences appear at two loops as double poles in $\e$, both in the leading and subleading terms. 

A comparison with the structure of IR divergences in ${\cal N}=4$ SYM amplitudes discloses a number of considerable differences.  

First of all, while in ${\cal N}=4$ SYM amplitudes divergences already appear at one loop, in three dimensions the first singularity is delayed at second order. Based on this observation, in \cite{CH, BLMPS1,BLMPS2} a comparison between the planar four--point amplitude in ${\cal N}=4$ SYM  at one loop and the same amplitude in ABJM at two loops has been discussed. A perfect identification between the two results, in particular for what concerns IR divergences, has been found upon rescaling $\e \to 2\e$ and formally identifying the mass scales. 
Instead, for finite $N$ the subleading contributions spoil this identification. 

To begin with, in ${\cal N}=4$ SYM double trace partial amplitudes appear already at one loop, while they are subleading in $\epsilon$ for the ABJM theory, at least up to two loops. Moreover, in the four dimensional case subleading contributions to the amplitude have milder IR divergences compared to the leading ones \cite{Naculich:2008ys}. In fact, the leading $\epsilon^{-2L}$ pole of a $L$--loop amplitude has been found to cancel in subleading contributions and the most subleading--in--color partial amplitude goes as $\epsilon^{-L}$.  More generally, it has been proved that $N^k$--subleading terms have at most $\epsilon^{-2L+k}$ poles.
Instead, for ABJM theory cancellation of leading poles does not occur at two loops and the leading and subleading partial amplitudes have the same leading singularity $1/\e^2$. 
This is basically due to the different color structures appearing in the two theories and can be better understood by constructing the two--loop operator which generates IR divergences in the ABJM theory, when applied to the tree level amplitude.  

We can define an abstract color space spanned by the basis of four traces (\ref{traces}) onto which projecting chiral amplitudes of the form $( (A^i)^{i_1}_{\ \bar{i}_1} (B_j)^{\bar{i}_2}_{\ i_2}  (A^k)^{i_3}_{\ \bar{i}_3} (B_l)^{\bar{i}_4}_{\ i_4})$ . In such a space the whole amplitude is thus represented as a four--vector. For instance, the tree level amplitude is proportional to $(1,-1,0,0)$.

Following what has been done in four dimensions \cite{Catani, Glover}, we define the operator $I^{(2)}(\epsilon)$ as a matrix acting on such a space and providing the IR divergences arising at second order coming from exchanges of two soft gluons between external legs
\begin{eqnarray}
\label{eq:matrix}
&& I^{(2)}(\epsilon) = -\frac{e^{-2 \gamma_E \epsilon} (8\pi)^{2\epsilon}}{(2\epsilon)^2}\,
\\&& 
\footnotesize
\left(\begin{array}{cccc}
(N^2-2)({\tt S}+{\tt T}) + 2 {\tt U} & 0 & N({\tt T}-{\tt U}) & N({\tt S}-{\tt U}) \\
0 & (N^2-2)({\tt S}+{\tt T}) + 2 {\tt U} & N({\tt T}-{\tt U}) & N({\tt S}-{\tt U}) \\
N({\tt S}-{\tt U}) & N({\tt S}-{\tt U}) & 2(N^2-1) {\tt S} + 2({\tt U} - {\tt T}) & 0 \\
N({\tt T}-{\tt U}) & N({\tt T}-{\tt U}) & 0 & 2(N^2-1) {\tt T} + 2({\tt U}-{\tt S})  
\end{array}\right) \nonumber
\end{eqnarray}
where we have defined
\begin{equation}
{\tt S} = (s/\mu^2)^{-2\epsilon} \, , \qquad {\tt T} = (t/\mu^2)^{-2\epsilon}\, , \qquad {\tt U} = (u/\mu^2)^{-2\epsilon}
\end{equation}
The action of such an operator on the tree level amplitude gives the structure of divergences for the complete four--point two--loop amplitude. In particular, when we apply it to the tree level vector $(1,-1,0,0)$, double trace contributions cancel. This stems for the absence of double trace contributions in the ABJM two--loop amplitude.

We note that the upper and lower $2 \times 2$ blocks on the right of matrix (\ref{eq:matrix}) are not required for our two--loop calculation. However, we have spelled them out for completeness: In principle, they might be required at higher orders if the IR divergences were to exponentiate in a similar manner to what happens in four dimensions. 

It is interesting to compare this matrix with the analogous ones in QCD \cite{Glover} and ${\cal N}=4$ SYM \cite{Naculich:2008ys} at one loop. Apart from the different dimensions obviously due to the different dimensions of the corresponding color spaces, they share the same configuration of leading IR divergences: While the leading--in--$N$ diagonal terms go like $1/\e^2$, the subleading--in--$N$ off--diagonal terms go like $1/\e$. However, the different structure of the tree--level amplitudes allows for the appearance of $1/\e$ divergent double trace contributions in four dimensions, which are not present in three dimensions.

\subsection{BLG amplitude}

BLG theory is the only model with $OSp(4|8)$ superconformal invariance in three dimensions. It can be realized as an ABJM theory with gauge group $SU(2)\times SU(2)$.
Therefore, we can use the previous results to get the complete two--loop amplitude ratio.

Even though the gauge group is actually $SU(2)\times SU(2)$, rather than $U(2)\times U(2)$ as would be for the ABJM theory, it turns out that this does not affect the color structure of the amplitude. 
Indeed, although extra terms from the subleading part of the gluon contractions appear in individual diagrams (with color factor up to $\sim N^{-2}$), all such contributions drop 
out and the final result turns out to be the same as the one of the ABJM case.
Therefore, setting $N=2$ in eq. (\ref{result}) the result reads
\begin{align}
\label{resultBLG}
{\cal M}_4^{BLG} &= \frac{1}{K^2} \biggl( \frac{-2\ (s/\m'^2)^{-2\e} -2\ (t/\m'^2)^{-2\e} - 2\ (u/\m'^2)^{-2\e}}{(2\e)^2} + \nonumber \\& \qquad \quad 
+ \log ^2 \frac{s}{t} + \log ^2 \frac{s}{u} + \log ^2 \frac{t}{u}
+  3 \pi^2 + 9 \log^2 2 \biggr)  + {\cal O}(\epsilon)
\end{align}
It is very interesting to observe how leading and subleading contributions in (\ref{result}) combine in order to give a result which is completely symmetric in any exchange of external labels.
This is manifest in the IR divergent piece and in the finite term.

Multiplying this by the tree level four--point superamplitude \cite{Huang:2010rn}, we obtain a two--loop superamplitude which is totally antisymmetric under any exchange of external labels. This is consistent with the fact that the theory possesses an underlying three algebra with a four--index structure constant $f^{abcd}$ which is totally antisymmetric. \\

\section{A superfield computation of the Sudakov form factor}\label{sec:formfactor}

The Sudakov form factor for the ABJ(M) theory in the planar limit has been evaluated up to two loops by Feynman diagrams \cite{Young:2013hda} and by unitarity cuts \cite{Brandhuber:2013gda}.
In this Section we exploit the previous results to provide an alternative evaluation of the Sudakov form factor based on a supergraph calculation and valid at any order in $N$.

In ordinary perturbation theory, the evaluation of form factors and scattering amplitudes are intimately connected whenever diagrams contributing to form factors can be obtained from diagrams contributing to amplitudes by simply collapsing free external matter legs into a bubble representing the operator insertion. 

This operation is particularly effective in superspace, given the peculiar structure of diagrams contributing to the four--point chiral amplitudes. In fact, since loop contributions always arise from corrections to the quartic superpotential vertex, it turns out that collapsing two free external legs in the supergraphs of Figs. \ref{fig:1set}, \ref{fig:np} we generate all the two--loop corrections to the form factor of a quadratic matter operator. As a consequence, the loop integrals appearing in the two computations are exactly the same. Only combinatorics and color factors in front of them are different. 

More precisely, for ABJ(M) theories we consider the following projection of the superfield form factor
\beq
\F(s) = \left \langle\, A_1(p_1)\, B_1(p_2)\, |\, \Tr (A_1\,  B_1)(p_1+p_2)\, |\, 0\, \right\rangle
\eeq
At one loop there is only one single diagram contributing, which comes from collapsing the one--loop diagram of the amplitude. As in the amplitude case  \cite{BLMPS1,BLMPS2}, the corresponding integral is ${\cal O}(\e)$, therefore negligible in three dimensions. 

At two loops, quantum corrections can be read from Figs. \ref{fig:1set}, \ref{fig:np} where we collapse two free external legs into the insertion of the operator $\Tr (A_1\,B_1)$. In this procedure we discard diagram $1(f)$ since it reduction simply does not exist, since it does not have two free external lines.  

A simple evaluation of the relevant color factors emerging from each graph leads to the following results (we still indicate $(p_1+p_2)^2 \equiv s$)
\begin{eqnarray} 
\F^{(a)} &=&  \left(\frac{4\pi}{K}\right)^2 \  (M-N)^2 \ { \cal D}_a(s) \nonumber \\
\F^{(b)} &=&  \frac14 \left(\frac{4\pi}{K}\right)^2 \  (M^2+N^2-4MN+2) {\cal D}_b(s) \nonumber \\
\F^{(c)} &=&  \frac14 \left(\frac{4\pi}{K}\right)^2 \left(M^2 +N^2 -8 MN + 6 \right)  \left({ \cal D}_b(s) - 2{\cal G}_{d}(p_1) - 2{\cal G}_{d}(p_2)\right) \nonumber \\
\F^{(d)} &=&  \ \ \left(\frac{4\pi}{K}\right)^2 \  (M^2+N^2-4MN+2) {\cal D}_d(s) \nonumber \\
\F^{(e)} &=& 2 \left(\frac{4\pi}{K}\right)^2 \ (2MN-2) {\cal D}_e(s) \nonumber  \\
\F^{(g)} &=& \frac12 \left(\frac{4\pi}{K}\right)^2 (M^2+N^2-8MN+6) \ \left({\cal G}_{d}(p_1)+{\cal G}_{d}(p_2)\right) \nonumber \\
\F^{(np)} &=&\ \ \left(\frac{4\pi}{K}\right)^2 \  (-2MN+2) {\cal D}_{np}(s)
\end{eqnarray}
where ${\cal D}$ and ${\cal G}_d$ integrals are given in eqs. (\ref{integrals1}, \ref{integral2}, \ref{np}). We note that also in ordinary Feynman diagram approach, as it happens in unitarity based calculations, a non--planar diagram ${\cal D}_{np}$ contributes to determine the final result also in the planar limit.   

First, setting $M=N$ and summing all the contributions, we find
\beq
\F_{ABJM}(s) = 2 \left( \frac{4 \pi}{K} \right)^2 \,  (N^2-1) \, \left( {\cal D}_d(s) - {\cal D}_{np}(s) \right)
\eeq
Inserting the results (\ref{integrals1}, \ref{np}) for the two integrals, we obtain the complete form factor at two loops for the ABJM theory 
\begin{equation}
\label{FF}
\F_{ABJM}(s) = 
\frac{(N^2-1)}{4 K^2} \left(-\frac{e^{-2 \gamma_E  \epsilon } (8 \pi \mu^2)^{2 \epsilon } s^{-2 \epsilon }}{\epsilon ^2}+\frac{2 \pi ^2}{3}+6 \log ^2 2 \right) + {\cal O}(\epsilon)
\end{equation}
The leading contribution in $N$ coincides with the result of \cite{Brandhuber:2013gda} under the identification $K = 4\pi k$ between the two Chern--Simons levels.
For finite $N$, expression (\ref{FF}) represents the complete non--planar result. Curiously, the subleading part combines in such a way that it is proportional to the leading one.

In the generalized unitarity approach the planar two--loop contribution to the Sudakov form factor turns out to be given in terms of a single crossed triangle integral $XT(s)$ (see eq. (4.14) in \cite{Brandhuber:2013gda}). Comparing that result with the present one an interesting relation is obtained among the integrals
\beq
XT(s) = 2 \left( {\cal D}_d(s) - {\cal D}_{np}(s) \right)
\eeq
More generally, for $M \neq N$, summing the previous contributions we obtain the complete form factor for the ABJ theory. In the planar limit, it reads
\begin{align}
\F_{ABJ}(s) &= \frac{1}{2 K^2} \, \left(\frac{e^{\gamma_E}\, s}{4\pi\mu^2} \right)^{-2 \epsilon} \, 
\left(-\frac{M N}{2 \epsilon ^2} - \log 2\, \frac{\left(M^2+N^2\right)}{2 \epsilon } + \right. \nonumber\\& \left. - \frac{1}{24} \pi ^2 \left(11 M^2 - 30 M N + 11 N^2\right) + \log ^2 2 \left(M^2+N^2\right)\right) + {\cal O}(\epsilon)
\end{align}
and agrees with the result of \cite{Young:2013hda}. \\

\section{Subleading contributions to the light--like Wilson loop}\label{sec:WL}

The four--point amplitude for ABJ(M) theories is MHV and in the planar limit it has been proved to match the light--like four--polygon Wilson loop up to two loops.  

The expectation value of the Wilson loop has been shown to vanish at one loop for any number of cusps \cite{HPW,BLMPRS} and calculated at two loops in the planar limit in \cite{HPW} for four cusps, and extended to $n$ cusps in \cite{Wiegandt}. Recently, the four--cusps planar computation has been refined in \cite{Bianchi:2013pva}, with a regularization scheme that preserves uniform transcendentality\footnote{This scheme also provides agreement between the perturbative computation of the expectation value of the $1/2$ BPS Wilson loop in ABJM and its exact result from localization \cite{BGLP}.}. The result can be written as $ \langle W_4 \rangle_{\rm ABJM} =  \langle W_4 \rangle_{\rm CS} +  \langle W_4 \rangle_{\rm matter}$, where in euclidean space
\bea
\label{eq:CS}
\langle W_4 \rangle_{\rm CS} &=& 1 -\left(\frac{N}{K}\right)^2  \frac{1}{4}\left[\log 2\, \sum_{i=1}^4\frac{(x_{i,i+2}^2 \, \pi e^{\g_E}  \mu^2 )^{2\epsilon}}{\epsilon} - 10 \zeta_2 + 8 \log^2 2 \right] + {\cal O}(K^{-3},\epsilon) \non \\
\eea
is the contribution from the pure Chern--Simons sector and 
\bea
\label{eq:matter}
\langle W_4 \rangle_{\rm matter} &=& 1 -\left(\frac{N}{K}\right)^2  \frac{1}{4}\left[\frac12  \sum_{i=1}^4\frac{(x_{i,i+2}^2 \, 4 \pi e^{\g_E} \mu^2 )^{2\epsilon}}{\epsilon^2} -   2\log{\frac{x_{13}^3}{x_{24}^2}} - \pi^2 \right] + {\cal O}(K^{-3},\epsilon) \non \\
\eea
is the contribution from the matter sector. Here, we have used the notation $x_{i,j} = x_i - x_j$, where $x_i$ label the polygon vertices. 

Summing the two contributions and rescaling the mass regulator as $\tilde \mu^2 = 8 \pi e^{\g_E} \mu^2$ the result reads 
\bea
\label{WLABJM} 
&& \langle W_4 \rangle_{\rm ABJM} =1 + \left( \frac{N}{K}\right)^2 \left[ - \frac{( { x_{13}^2 \, \tilde \mu}^2)^{2\epsilon}}{(2\epsilon)^2} -
\frac{({x_{24}^2 \, \tilde \mu}^2)^{2\epsilon}}{(2\epsilon)^2}  + \frac12  \log^2\left(\frac{x_{13}^2}{x_{24}^2}\right) \right.
\nonumber\\
&& \hspace{6cm} + \frac{2 \pi^2}{3} + 3 \log^2 2 \Big] + {\cal O}(K^{-3},\epsilon)
\eea
In fact, at this order it coincides with the result for the amplitude, eq. (\ref{planar}), both in its divergent and in its non--constant and constant finite parts if we formally identify $s = x_{13}^2$, $t = x_{24}^2$ and send $\e \to -\e$ and $\mu^{\prime} \to 1/\tilde \mu$. 

Having computed the complete two--loop amplitude for finite $N$ it is then interesting to investigate the subleading contributions to the four--polygon Wilson loop. 

\subsection{Non--planar diagrams}

\FIGURE{ 
    \centering
    \includegraphics[width=.2\textwidth]{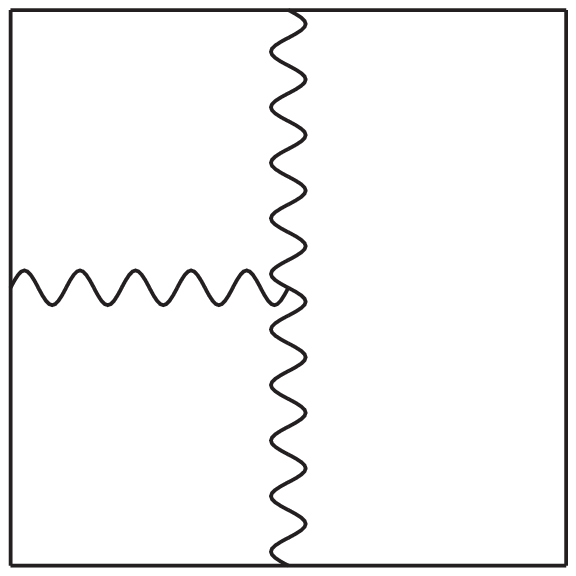}
    \caption{Vertex diagram.}
    \label{fig:vertex}}

It is easy to realize that subleading contributions arise at two loops only in the pure Chern--Simons sector. Therefore, we can restrict the analysis to this sector.

We recall that in the planar limit contributions in eq. (\ref{eq:CS}) come from the three--vertex graph\footnote{Light--like Wilson loops are defined in terms of the gauge vector components $A_\mu, \hat{A}_\mu$ of the two gauge superfields $V, \hat{V}$. Therefore, their expectation value is computed perturbatively by ordinary Feynman diagrams, not superdiagrams.} in Fig. \ref{fig:vertex} and ladder graphs where the exchange of two non-crossing gauge lines appears. All contributions have a leading color factor $N^2$. 

Subleading in $N$ terms originate from two sources: The subleading piece of the vertex diagram of Fig. \ref{fig:vertex}, whose complete color factor is $(N^2-1)$, and new genuinely non--planar diagrams.

In principle, there are five new potential graphs with non--planar configurations of two gauge lines. However, because of the antisymmetry of the $\epsilon$ tensor carried by gluon propagators (see eq. (\ref{propcomponents})), only two of them give non--vanishing contributions. Defining $ z_i^\mu(s) = x_i^\mu + x_{i+1, i}^\mu s$ where $0 \leq s \leq 1$ is the affine parameter of the wedge $i$, the corresponding integrals can be written as
\begin{equation}\label{eq:nonplanar1}
I_{1}^{(np)} = \raisebox{-0.6cm}{\includegraphics[width=0.1\textwidth]{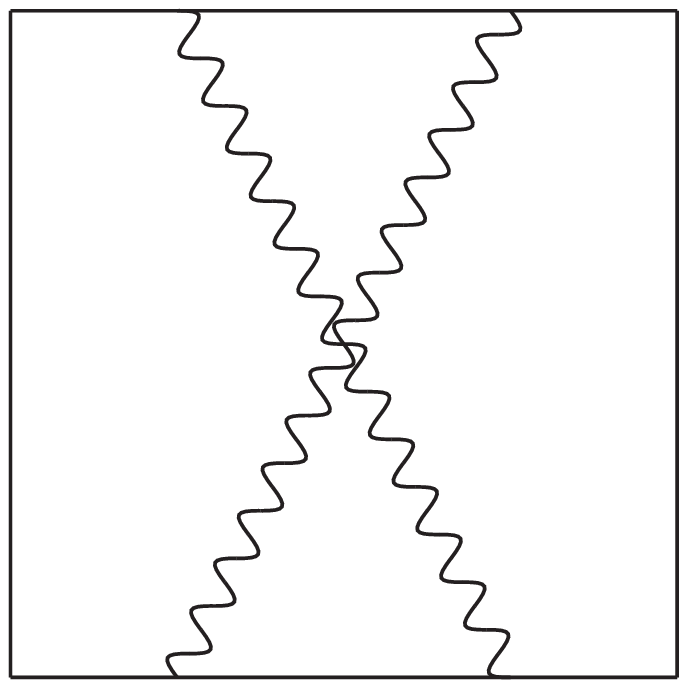}} = \int [ds]_4 \frac{\varepsilon(\dot z_1(s_1), \dot z_3(s_3), z_{13}) \varepsilon(\dot z_1(s_2), \dot z_3(s_4), z_{24})}{\left(\bar s_1 \bar s_3 x_{13}^2 + s_1 s_3 x_{24}^2\right)^{3/2} \left(\bar s_2 \bar s_4 x_{13}^2 + s_2 s_4 x_{24}^2\right)^{3/2}}
\end{equation}
and
\begin{equation}\label{eq:nonplanar2}
I_{2}^{(np)} = \raisebox{-0.6cm}{\includegraphics[width=0.1\textwidth]{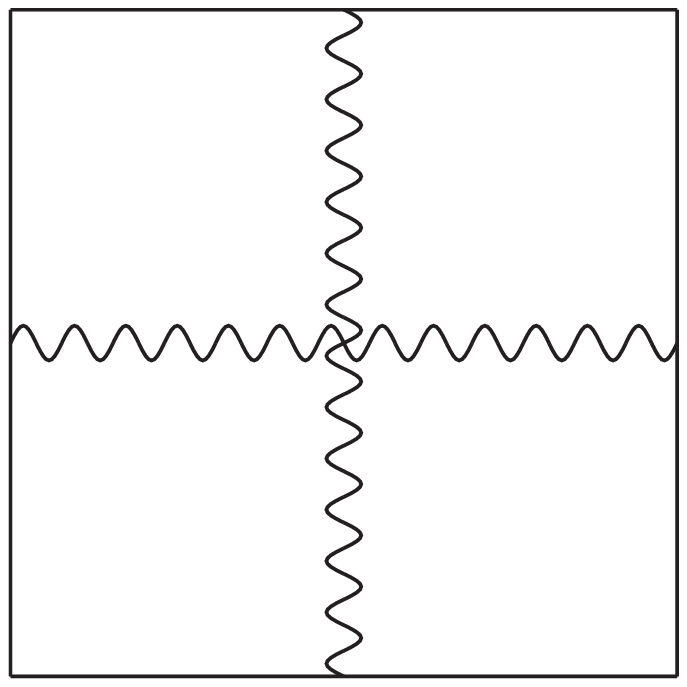}} = \int [{\tilde {ds}}]_4 \frac{\varepsilon(\dot z_1(s_1), \dot z_3(s_3), z_{13}) \varepsilon(\dot z_2(s_2), \dot z_4(s_4), z_{24})}{\left(\bar s_1 \bar s_3 x_{13}^2 + s_1 s_3 x_{24}^2\right)^{3/2} \left(\bar s_2 \bar s_4 x_{24}^2 + s_2 s_4 x_{13}^2\right)^{3/2}}
\end{equation}
Here we have defined 
\begin{equation}
\int [ds]_4 = \int_0^1 ds_1 \int_0^{s_1}  ds_2 \int_0^1 ds_3 \int_0^{s_3} ds_4 \qquad \int [{\tilde {ds}}]_4 = \int_0^1 ds_1 ds_2 ds_3 ds_4 \nonumber
\end{equation}
and $\varepsilon(a,b,c) =  \varepsilon_{\mu \nu \rho} a^\mu b^\nu c^\rho$. 
The integrations over the affine parameters turn out to be finite, so that no regularization is required.

The two integrals are related to the planar ladder diagram by the following identity 
\begin{equation}\label{eq:relation}
I^{(ladder)}+I_1^{(np)} = -\frac12\, I_2^{(np)}
\end{equation}
This can be understood by observing that the left hand side turns out to be the integral $ \int [ds]_4$ of an expression which is  symmetric under exchanges $s_1 \leftrightarrow s_2$ and  $s_3 \leftrightarrow s_4$. Therefore, the integration region can be also symmetrized as $\frac{1}{4} \int [{\tilde {ds}}]_4$. Moreover, elaborating the numerators and performing simple changes of integration variables, it can be shown that the integrand can be reduced to be the same as the one in $ I_2^{(np)}$. 

Exploiting identity (\ref{eq:relation}) it is easy to realize that the non--planar combination entering the computation can be rewritten solely in terms of planar contributions, as graphically illustrated in Fig. \ref{fig:relation}.
\FIGURE{
\centering
\includegraphics[width=1\textwidth]{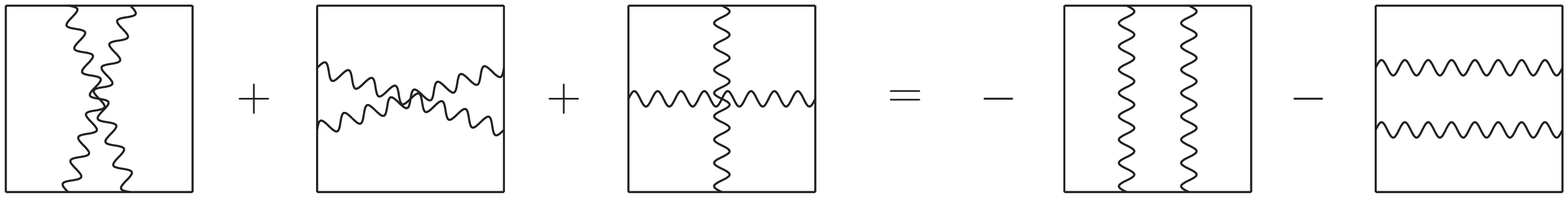}\caption{Graphical relation between integrals.}\label{fig:relation}
}
The sum of the non--planar integrals is then exactly equal to the sum of planar ladder diagrams appearing in the planar part, but this time with subleading color factor $(-1)$. Therefore, the effect of the new non--planar contributions is simply to modify the original color factor $N^2 \to (N^2 - 1)$ in front of the ladder diagrams.
This combines nicely with the subleading color factor $(N^2-1)$ in front of the vertex diagram, so that the final result at finite $N$ is exactly the same as the planar one (\ref{eq:CS}) except for a change in the overall color factor
\begin{align}\label{eq:CSWL}
\langle W_4 \rangle_{\rm CS} &= 1 - \frac{N^2-1}{4K^2} \left[\log{2} \sum_{i=1}^4\frac{(x_{i,i+2}^2 \,\pi e^{\g_E} \mu^2 )^{2\epsilon}}{\epsilon} - 10 \zeta_2 + 8 \log^2{2} + {\cal O}(\e) \right] + {\cal O}(K^{-3})
\end{align}
The result is manifestly maximally transcendental.
 
The expectation value for the the ABJ(M) theories can be easily obtained by adding the contribution (\ref{eq:matter}). It is straightforward to see that subleading contributions break the duality between the amplitude and the corresponding Wilson loop, as expected.

\subsection{Interpretation from non--abelian exponentiation theorem}

The simple result highlighted above, in particular the factorization of the complete color factor is not a coincidence, since it has a nice explanation in terms of the so--called non--abelian exponentiation theorem of Wilson loops \cite{Gatheral:1983cz,Frenkel:1984pz}.

This theorem states that the perturbative computation of the expectation value of a Wilson loop can be rearranged as an exponential where only certain Feynman diagrams appear and with a color factor different from the ordinary one.

More precisely, classifying color graphs in terms of webs of gluon lines, the non--abelian exponentiation theorem states that the Wilson loop expectation value takes the form
\begin{equation}
\langle W \rangle = \exp \left( \sum_{L=1}^{\infty} c_L\, w_L \right)
\end{equation}
where in the exponent the sum is over all loops and at a given order $w_L$ is expressed as a sum of ``single webs".  Using Jacobi identities, such webs can be decomposed into the product of connected webs, 
as explained in \cite{Frenkel:1984pz}. The corresponding color factor $c_L$ is then the one associated to the so--called ``color connected diagram" appearing in the decomposition, that is the color diagram containing only one connected web.  

We can evaluate the Wilson loop for pure Chern--Simons theory by using this prescription. At one loop there is just one diagram and its total contribution vanishes $w_1=0$.
Therefore, at two loops where in principle the result would have been given by the combination of one--loop and two--loop webs $w_1$ and $w_2$ 
\begin{equation}
\langle W_4^{CS} \rangle^{(2)} = \frac12 (c_1 w_1)^2 + c_2 w_2 = c_2 w_2
\end{equation}
we have just to take into account $w_2$.

There are three single web diagrams at two loops: the planar vertex integral in Fig. \ref{fig:vertex}, and the non--planar contributions (\ref{eq:nonplanar1}) and (\ref{eq:nonplanar2}).
 
The vertex diagram is already a color connected web, therefore its color factor is $c_2 = (N^2-1)$. Decomposing the two non--planar diagrams into products of connected webs as in Fig. \ref{fig:decomposition} their color factor is the one associated to the color connected diagram, that is the diagram containing the three--gluon vertex. Therefore, it is still $c_2 = (N^2 - 1)$ as for the planar contribution.  

\FIGURE{
\centering
\includegraphics[width=0.6\textwidth]{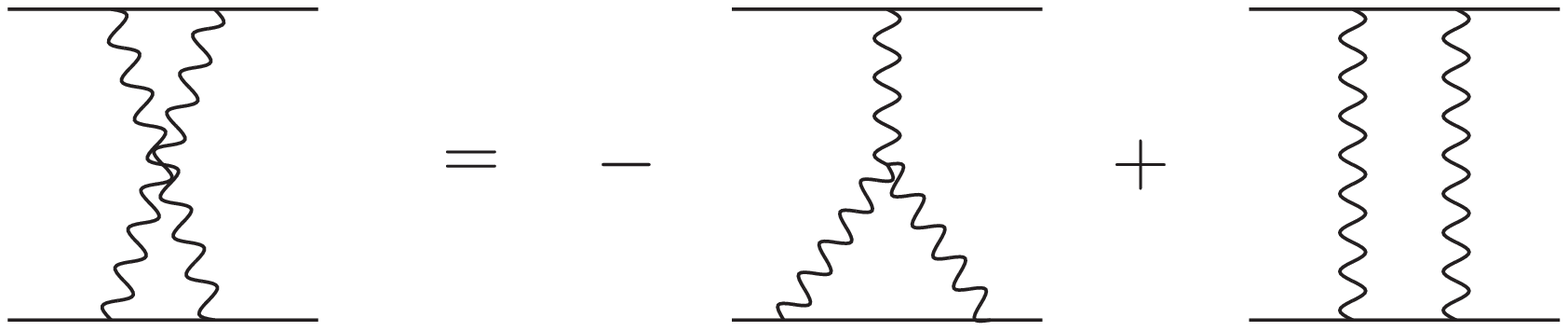}\caption{Web decomposition of a non--planar diagram into one ``color connected" web and two connected webs. The color factor $c_L$ is the one associated to the first graph.}\label{fig:decomposition}
}

In principle, this is enough to get the complete answer for the two--loop Wilson loop. In fact, we already know the coefficient of the leading $N^2$ term in the Wilson loop expectation value from a previous computation. From information obtained from web exponentiation it follows that the total result must be simply $(N^2-1)$ times the planar result. This is indeed what we have found by performing an explicit calculation.

We could have computed the complete Wilson loop expectation value using the non--abelian exponentiation theorem from the very beginning. Along with the vertex diagram, we should consider the non--planar diagrams. As stressed above, the latter are not color connected webs and have to be decomposed as in Fig. \ref{fig:decomposition}.
In such a decomposition we observe that their color connected component comes with a minus sign.
This means that the non--planar diagrams enter the computation of $w_2$ with the color factor $-(N^2-1)$. Taking into account the graphical equation of Fig. \ref{fig:relation}, relating the non--planar integrals with the ladder ones, we see that the final combination of non--planar diagrams is equivalent to the ladder diagram contribution of the planar result, but with color factor $(N^2-1)$.
This is in agreement with the computation of the previous Section. \\

\section{Conclusions}

In this paper we have tackled the problem of evaluating physical observables in $U(M) \times U(N)$ ABJ(M) theories for finite $M, \, N$. In particular, we have focused on the four--point scattering amplitude and, related to it, on the Sudakov form factor and the four--cusp light--like Wilson loop. We have evaluated them up to two loops. Although the most interesting features, like dualities and extra symmetries, are expected to arise in the planar limit, the evaluation of quantities for finite ranks of the gauge groups gives useful information about the complete structure of IR (UV) divergences, also in connection with supergravity amplitudes. Moreover, from results at finite $N$ we can read the quantum corrections to observables in the BLG model. 

The complete two--loop four--point amplitude that we have obtained for ABJM possesses interesting properties. First of all, at least at two loops double trace partial amplitudes cancel completely in the final result. Moreover, subleading--in--$N$ contributions share with the planar part the same degree of leading IR singularities. These are novelties if compared to the ${\cal N}=4$ SYM case. In fact, given the different color structure of the theory, in four dimensions double trace divergent terms appear already at one loop. Furthermore, non--trivial cancellations of the leading poles in the non--planar part of the amplitudes occur, which do not seem to have an analogue in the three dimensional ABJM model. 

In ${\cal N}=4$ SYM, IR divergences associated to the most subleading--in--$N$ terms have been conjectured to exponentiate and to give rise to the IR structure of the corresponding 
${\cal N}=8$ supergravity amplitudes obtained by the double--copy prescription \cite{Naculich:2013xa}. It would be very interesting to investigate whether the IR divergent contributions exponentiate also for ABJM. This would necessarily require the evaluation of the IR divergent part of the amplitude, at least at the next non--trivial order, that is four loops. 

The connection of BLG amplitudes with those of the corresponding ${\cal N}=16$ supergravity via the color/kinematics duality of the gauge theory and the double--copy property of gravity is still to be widely investigated.  
As discussed in \cite{Bargheer:2012gv}, BLG amplitudes can be written in such a way that BCJ--like relations \cite{BCJ} hold and in principle can be used to construct supergravity amplitudes as double copies of the gauge ones. In particular, this first requires expressing the whole amplitude in terms of a suitable basis of color factors related by Jacobi identities. At tree level the color/kinematics duality states that it is possible to rearrange the amplitude in such a way that the kinematic coefficients associated to those color structures obey a corresponding Jacobi identity. 
At loop level unitarity allows to reconstruct loop integrals from tree level amplitudes. Applying the aforementioned BCJ relations to these yields a set of constraints which the integrands have to satisfy, mixing in particular planar and non--planar contributions.
A convenient way to begin realizing this program in ABJM would be to re--derive our result for the complete four--point amplitude by using a unitarity based approach. 
This is currently under investigation.

The non--planar contributions to ABJ(M) observables do not spoil the uniform transcendentality of the planar results. This is analogous to what has been observed for ${\cal N}=4$ SYM amplitudes \cite{Naculich:2008ys}. \\

\section{Acknowledgements}

MB thanks Lorenzo Bianchi for very useful discussions.
We also thank Donovan Young for correspondence.  
The work of MB has been supported by the Volkswagen-Foundation.
The work of ML has been supported by the research project CONICET PIP0396.
The work of SP has been supported in part by INFN, MIUR--PRIN contract 2009--KHZKRX and MPNS--COST Action MP1210 "The String Theory Universe".

\vfill
\newpage

\appendix

\section{The ABJ(M) theory in ${\cal N}=2$ superspace}

In three dimensional ${\cal N}=2$ superspace \cite{Klebanov}, the field content of the $U(M) \times U(N)$ ABJ(M) theories is given in terms of two vector multiplets $(V,\hat{V})$
in the adjoint representation of the first and the second group respectively,  
coupled to four chiral multiplets $(A^i)^a_{\ \bar{a}}$ and $(B_i)^{\bar{a}}_{\ a}$ carrying a fundamental index $i=1,2$ of a global $SU(2)_A \times SU(2)_B$ and in the (anti)bifundamental representations of the gauge group ( $a$ and $\bar{a}$ are indices of the fundamental representation of the first and the second gauge groups, respectively).

In euclidean superspace with the effective action defined as $e^\G = \int e^S$ the action reads
\begin{equation}
 {\cal S} = {\cal S}_{\mathrm{CS}} + {\cal
    S}_{\mathrm{mat}}
  \label{eqn:action}
\end{equation}
with
\begin{align}
  \label{action}
& {\cal S}_{\mathrm{CS}}
=  \frac{K}{4\pi} \, \int d^3x\,d^4\theta \int_0^1 dt\: \Big\{  \Tr \Big[
V\, \Db^\a \left( e^{-t V}\, D_\a\, e^{t V} \right) \Big]
-\Tr \Big[ \hat{V}\, \Db^\a \left( e^{-t \hat{V}} D_\a
e^{t\, \hat{V}}\, \right) \Big]   \Big\} \nonumber \\
& {\cal S}_{\mathrm{mat}} = \int d^3x\,d^4\theta\: \Tr \left( \bar{A}_i\,
e^V\, A^i\, e^{- \hat{V}} + \bar{B}^i\, e^{\hat V}\, B_i\,
e^{-V} \right)\nonumber \\
&  +\frac{2\pi i}{K}\int d^3x\, d^2\theta\, \epsilon_{ik}\,\epsilon^{jl}\,
\Tr\,\left(A^i\, B_j\, A^k\, B_l\right)+\frac{2\pi i}{K}\int d^3x\, d^2\bar\theta\,
\epsilon^{ik}\,\epsilon_{jl}\, \Tr\,\left(\bar A_i\, \bar B^j\, \bar A_k\, \bar B^l\right)
\end{align}
Here $K$  is the Chern--Simons level. It must be an integer, as required by gauge invariance.
For superspace conventions we refer to \cite{BLMPS1,BLMPS2} and, in particular, to Appendix B in \cite{Bianchi:2012cq}. 

In scattering amplitudes the external particles satisfy the free equations of motion
\beq
\label{onshell}
D^2 A^i = D^2 B_i = 0 \qquad , \qquad \bar{D}^2 \bar{A}_i = \bar{D}^2 \bar{B}^i = 0
\eeq
The quantization of the theory can be easily carried out in superspace.
After performing gauge fixing (for details, see for instance \cite{BPS}), in Landau gauge the super--vector propagators are
\bea
  \langle V^a_{\, b}(1) \, V^c_{\, d}(2) \rangle
  &=&   \frac{4\pi}{K} \, \frac{1}{p^2} \,  \, \delta^a_d \, \d^c_b \times \Db^\a D_\a \, \delta^4(\th_1-\th_2) \nonumber \\
  \langle \hat V^{\bar{a}}_{\bar{b}} (1) \, \hat V^{\bar{c}}_{\bar{d}}(2) \rangle &=& -
  \frac{4\pi}{K} \,  \frac{1}{p^2} \, \,   \delta^{\bar{a}}_{\bar{d}} \, \d^{\bar{c}}_{\bar{b}} \times \Db^\a D_\a  \, \delta^4(\th_1-\th_2)
  \label{gaugeprop}
\eea
whereas the matter propagators read
\begin{eqnarray}
  &&\langle \bar A^{\bar{a}}_{\ a}(1) \, A^b_{\ \bar{b}}(2) \rangle
  = \frac{1}{p^2} \,\, \delta^{\bar{a}}_{\ \bar{b}} \, \delta^{\ b}_{a} \times  D^2 \bar{D}^2 \, \delta^4(\th_1 - \th_2)
 \nonumber \\
  &&  \langle \bar B^a_{\ \bar{a}}(1) \, B^{\bar{b}}_{\ b}(2) \rangle =
   \frac{1}{p^2} \,\, \delta^a_{\ b} \, \delta^{\ \bar{b}}_{\bar{a}} \times D^2 \bar{D}^2 \, \delta^4(\th_1 - \th_2)
\label{scalarprop}
\end{eqnarray}
Going to components, in $3-2\e$ dimensions the propagators of the gauge fields $A_\mu, \hat{A}_\mu$ are
\bea
\label{propcomponents}
 \langle (A_\mu)^a_{\, b}(x) \, (A_\nu)^c_{\, d}(y) \rangle
  &=&   \left( \frac{2\pi i}{K} \right) \frac{\G(\frac32-\e)}{2\pi^{\frac32 -\e}} \varepsilon_{\mu\nu\rho} \frac{(x-y)^\rho}{[(x-y)^2]^{\frac32 -\e} } \, \, \delta^a_d \, \d^c_b 
\non \\
\langle ({\hat A}_\mu)^{\bar{a}}_{\bar{b}} (x) \, ({\hat A}_\nu)^{\bar{c}}_{\bar{d}}(y) \rangle &=&   - \left( \frac{2\pi i}{K} \right) \frac{\G(\frac32-\e)}{2\pi^{\frac32 -\e}} \varepsilon_{\mu\nu\rho} \frac{(x-y)^\rho}{[(x-y)^2]^{\frac32 -\e} } \, \,   \delta^{\bar{a}}_{\bar{d}} \, \d^{\bar{c}}_{\bar{b}} 
\eea

The vertices employed in our two--loop calculations can be easily read from the action (\ref{action}) and they are given by
\begin{align}
  \label{vertices}
 & \int d^3x\,d^4\theta\, \left[ \Tr ( \bar{A}_i V A^i) -  \Tr ( B_i V \bar{B}^i )
    + \Tr (  \bar{B}^i \hat V B_i ) -  \Tr ( A^i \hat{V}   \bar{A}_i ) + \right.
    \non \\
    &  +  \frac12\, \Tr ( \bar {A}_i \{ V,V\} A^i) 
    + \frac{1}{2}\, \Tr ( B_i \{V,V\} \bar B ^i ) + \frac12 \Tr ( A_i \{\hat{V},\hat{V}\} \bar A^i) +  \non \\ 
    & \qquad \left. + \frac{1}{2} \Tr ( \bar{B}_i \{\hat{V},\hat{V}\} B^i )  - \Tr (  \bar{B}^i {\hat V} B_i V ) -  \Tr ( A^i {\hat{V}}   \bar{A}_i V ) \right] +
    \non \\
& \qquad + \frac{4\pi i}{K} \int d^3x\,d^2\theta\,
    \, \Big[ \Tr (A^1 B_1 A^2 B_2) -  \Tr (A^1 B_2 A^2 B_1)\Big] ~+~ {\rm h.c.}
\end{align}

\section{Non--planar integral}\label{app:integral}

We compute the following integral (we drop the $(\mu^2)^{2\epsilon}$ factor for convenience)
\begin{equation}
{\cal D}_{np}(s) =  -\int \frac{d^d k}{(2\pi)^d} \frac{d^d l}{(2\pi)^d} \frac{ \Tr ( (k+l)\ k\ l \ (k+l)\ p_4\ p_3 )} {k^2 (k+l-p_3)^2 (k+p_4)^2 (l-p_3)^2 \ (k+l+p_4)^2 \ l^2}
\end{equation}
which emerges as non--planar contribution to the four--point amplitude. Along the calculation we will always make use of the on--shell conditions $p_i^2=0$.

We begin by making Feynman combining of $1/l^2$ and $1/(l-p_3)^2$ propagators
\begin{equation}
-\int \frac{d^dk}{(2\pi)^d}\,\frac{d^dl}{(2\pi)^d}\, \frac{\Tr(p_4 p_3 (k+l) k l (k+l))}{k^2 (k+p_4)^2 (k+l+p_4)^2 (k+l-p_3)^2} \int_0^1 d\alpha_2 \frac{1}{[(l-\alpha_2 p_3)^2]^2}
\end{equation}
Performing the change of variables $l\rightarrow r-k$ and elaborating the numerator with simple algebra we can write the integrand as the sum of two terms
\begin{equation}
\label{twopieces}
\int \frac{d^dk}{(2\pi)^d}\,\frac{d^dr}{(2\pi)^d}\, \frac{r^2\, [ \Tr(p_4p_3rk) - k^2 s]}{k^2 (k+p_4)^2 (r+p_4)^2 (r-p_3)^2} \int_0^1 d\alpha_2 \frac{1}{[(r-k-\alpha_2 p_3)^2]^2}
\end{equation}
We are going to analyze the two pieces separately.

\subsection{Integral 1)}

In the first term we first concentrate on the $k$--integration and Feynman parametrize the $1/k^2$ and $1/(k+p_4)^2$ propagators. Performing a harmless shift $k \to k - \a_1 p_4$
we end up with  
\begin{align}
\int \frac{d^dr}{(2\pi)^d}\, \frac{1}{(r+p_4)^2 (r-p_3)^2} \, \int_0^1 d\alpha_1 \int_0^1 d\alpha_2
\int \frac{d^dk}{(2\pi)^d}\,\frac{r^2\, \Tr(p_4p_3rk)}{(k^2)^2 [(k-r-\alpha_1 p_4+\alpha_2 p_3)^2]^2} 
\end{align}
where the $k$--integration can be immediately performed, being a vector bubble integral, leading to
\begin{equation}\label{eq:int1}
\frac12\, \int \frac{d^dr}{(2\pi)^d}\, \int_0^1 d\alpha_1 \int_0^1 d\alpha_2 \,
\frac{(r^2)^2\, s - \alpha_2\, s\, r^2\, 2\, p_3\cdot r}{(P^2)^{4-d/2}(r+p_4)^2 (r-p_3)^2} \, G[2,2]
\end{equation}
Here we have defined
\begin{equation}
P^2 = (\alpha_1 p_4 - \alpha_2 p_3 + r)^2
\end{equation}
and
\begin{equation}
\label{formula}
G[a,b] = \frac{\Gamma(a+b-d/2)\Gamma(d/2-a)\Gamma(d/2-b)}{(4\pi)^{d/2}\Gamma(a)\Gamma(b)\Gamma(d-a-b)}
\end{equation}
Completing the squares in the numerator of (\ref{eq:int1}) we obtain the sum of two scalar integrals
\begin{equation}
\frac12\, \int \frac{d^dr}{(2\pi)^d}\, \int_0^1 d\alpha_1\int_0^1 d\alpha_2 \, 
\frac{\bar \alpha_2 (r^2)^2\, s + \alpha_2\, s\, r^2\, (r-p_3)^2}{(P^2)^{4-d/2}(r+p_4)^2 (r-p_3)^2} \, G[2,2]
\end{equation}
where we have defined $\bar{\alpha}=1-\alpha$.

The second integral is very easy to compute. Setting $d=3-2\e$ and expanding the result in powers of the dimensional regulator, we obtain
\begin{equation}
\frac12\, G[2,2]\, \int \frac{d^dr}{(2\pi)^d}\, \int_0^1 d\alpha_1\, \int_0^1 d\alpha_2 
\frac{\alpha_2\, s\, r^2}{(P^2)^{4-d/2}(r+p_4)^2} = \frac{1}{64 \pi^2} + {\cal O}(\epsilon)
\end{equation}
The first integral requires a little bit more of effort.
Using Mellin--Barnes representation allows to easily evaluate the $\a_i$ integrals. After a shift $(r^2)^2 \rightarrow (r^2)^{2-\delta}$ we obtain
\begin{align}
& \frac{2^{4 \epsilon} \pi ^{2 \epsilon}}{128 \pi^3}\int \frac{dudv}{(2\pi i)^2}\, (-1)^v s^{-\delta -2 \epsilon } \Gamma (-u) \Gamma (-v) \Gamma (-w) \Gamma (w+1) \Gamma (-u-w-\delta -2 \epsilon )
\nonumber\\ & 
\Gamma \left(-\epsilon -\frac{1}{2}\right)^2 \Gamma (u+w+1) \Gamma (v+\delta -2) \Gamma (v+w+1) \Gamma \left(-u-\delta -\epsilon +\frac{3}{2}\right) 
\nonumber\\ &
\frac{ \Gamma (-v-w-\delta -2 \epsilon ) \Gamma (-u-v-w-\delta -2 \epsilon +1) \Gamma (u+v+w+\delta +2 \epsilon +1)}{\Gamma (\delta -2) \Gamma (-2 \epsilon -1) \Gamma \left(-\delta -3 \epsilon +\frac{1}{2}\right) \Gamma (-u-\delta -2 \epsilon +1) \Gamma (-u-\delta -2 \epsilon +2)}
\end{align}
Expanding in $\delta$ and $\epsilon$ up to order zero terms one gets two remaining one--fold integrals 
\begin{equation}
\int \frac{du}{2\pi i} \Gamma(3/2-u) \Gamma(u) \left( \Gamma(-1+u)^{**} \Gamma(1-u) -2 \Gamma(u)^* \Gamma(-u) \right)
\end{equation}
where asterisks denote how many of the first right (left) poles of the $\Gamma$ functions have to be considered left (right), according to the notation of \cite{Smirnov}.
Such Barnes integrals can be solved by lemmas (D.12) and (D.37) of \cite{Smirnov}.
Summing the contributions gives
\begin{equation}
\label{int1}
\frac{1}{64 \pi ^2} \left[\frac{(16\pi)^{2\epsilon} s^{-2 \epsilon} e^{2 \epsilon (1-\gamma_E)}}{(2\epsilon)^2}-\frac{\pi ^2}{24}-\frac32 - 4 \log 2\, (1+\log 2) \right]
\end{equation}

\subsection{Integral 2)}

We now consider the second piece in eq. (\ref{twopieces}) with the shift $k \to k -p_4$
\begin{equation}
\int \frac{d^dr}{(2\pi)^d}\, \frac{-r^2\, s}{(r+p_4)^2 (r-p_3)^2} \int_0^1 d\alpha_2 \, 
\int \frac{d^dk}{(2\pi)^d}\,\frac{1}{k^2[(k-r-p_4+\alpha_2 p_3)^2]^2}
\end{equation}
and perform the bubble $k$--integral  
\begin{equation}
\int_0^1 d\alpha_2 \, 
G[1,2]\, \int \frac{d^dr}{(2\pi)^d}\, \frac{-r^2\, s}{(r+p_4)^2 (r-p_3)^2 [(r+p_4-\alpha_2 p_3)^2]^{3/2+\epsilon}}
\end{equation}
Shifting $r^2 \to (r^2)^{1-\delta}$ and using Mellin--Barnes representation, for $d=3-2\epsilon$ we have
\begin{align}
&-\frac{s^{-\delta -2 \epsilon}}{(4 \pi )^{3-2 \epsilon}}\, \frac{\Gamma \left(-\epsilon -\frac{1}{2}\right) \Gamma \left(\frac{1}{2}-\epsilon \right)}{\Gamma (\delta -1) \Gamma (-2 \epsilon ) \Gamma \left(-\delta -3 \epsilon +\frac{1}{2}\right)}\, \int_{-i \infty}^{+i\infty} \frac{du\,dv}{(2\pi i)^2}\,
(-1)^{v}\, 
\nonumber\\&
\Gamma (-u) \Gamma (-v) \Gamma (u+1) \Gamma (v+1)  \Gamma (v+\delta -1) \Gamma (-v-\delta -2 \epsilon ) 
\nonumber\\& 
\Gamma (-u-v-\delta -2 \epsilon )
\Gamma \left(-u-\delta -\epsilon +\tfrac{1}{2}\right) \frac{\Gamma (u+v+\delta +2 \epsilon +1)}{\Gamma (-u-\delta -2 \epsilon +1)}
\end{align}
Now, selecting poles that give an order $\delta^0$ result leads to a one--fold Mellin--Barnes integral, which can be expanded in $\epsilon$. The one--fold integral vanishes identically, leaving
\begin{equation}
\label{int2}
-\frac{1}{64 \pi ^2}\left( \frac{s^{-2\epsilon}e^{-2\epsilon \gamma_E} (4\pi )^{2\epsilon}}{2 \epsilon} -1-2 \log 2 \right)
\end{equation}

\subsection{Sum}

Summing the two contributions (\ref{int1}) and (\ref{int2}) it is interesting to observe that all terms of lower transcendentality cancel, leaving 
\begin{equation}
{\cal D}_{np}(s) = \frac{e^{-2\epsilon \gamma_E} (16 \pi)^{2 \epsilon} s^{-2 \epsilon}}{64 \pi^2} \left(\frac{1}{(2\epsilon)^2}-\frac{\pi ^2}{24}-4 \log ^2 2\right)
\end{equation}

\end{document}